\documentclass[conference]{IEEEtran}

\usepackage{amsmath}
\usepackage[siunitx]{circuitikz}
\usepackage[utf8]{inputenc}
\usepackage{biblatex}
\usepackage{siunitx}
\usepackage[hyphenbreaks]{breakurl}
\usepackage[breaklinks=true]{hyperref}
\bibliography{citation.bib}

\IEEEoverridecommandlockouts

\usetikzlibrary{arrows.meta,
                calc, chains,
                quotes,
                positioning,
                shapes.geometric}

\tikzstyle{startstop} = [rectangle, rounded corners, minimum width=3cm, minimum height=1cm,text centered, draw=black]
\tikzstyle{io} = [trapezium, trapezium left angle=70, trapezium right angle=110, minimum width=3cm, minimum height=0.8cm, text width=5em, text centered, draw=black]
\tikzstyle{process} = [rectangle, minimum width=3cm, minimum height=1cm, text centered, text width=5em, draw=black]
\tikzstyle{decision} = [diamond, aspect=2, minimum width=3cm, text width=3em, minimum height=1cm, text badly centered, draw=black]
\tikzstyle{arrow} = [thick,->,>=stealth]

\begin{document}

\title{Injecting Software Vulnerabilities with Voltage Glitching}

\author{\IEEEauthorblockN{Yifan Lu
\thanks{This work was not supported and do not represent the approval or rights of any third parties.}}
\IEEEauthorblockA{me@yifanlu.com}
}

\maketitle

\begin{abstract}

We show how voltage glitching can cause timing violations in CMOS behavior. Then we attack a real, security hardened, consumer device to gain code execution and dump the secure boot ROM.

\end{abstract}

\section{INTRODUCTION}

Glitching, or fault injection, has been used for over a decade \cite{ARKM} to attack software running on secure execution environments. Due to the upward trend in pricing in the software exploit market \cite{scip} and the increased hardening of security in consumer devices, there has been a rise in popularity of injecting faults to gain control of a device. Fault injections can be used to cause a malfunction in the target's system-on-chip (SoC) and, when the malfunction is controlled properly, can be used by an attacker to take full control of the device.

Voltage glitching is a specific kind of fault injection and is attractive because is it inexpensive to set up and is widely applicable to most chips \cite{6178001}. \textit{Crowbar voltage glitching} was introduced by O'Flynn \cite{OFlynn2016FaultIU} and implemented in the ChipWhisperer open hardware platform to bring these attacks to the mainstream. It works by abusing the capacitance ringing effect caused by introducing a crowbar circuit into the existing system. The ringing causes faults that can be exploited.

\subsection{Background}

We looked at the prior attempts at modeling voltage fault injections and ordered them in terms of abstraction (see table \ref{table:related}). Most recently, in a paper by Timmers, Spruyt, and Witteman \cite{TSW}, they created an architectural model of fault injection for ARM devices. Their model considers instruction corruption due to bit-flips caused by the fault. Their model is applicable to many kinds of fault injection.

Another paper we drew inspiration comes from Zussa, Dutertre, Clédière and Tria \cite{ZDCT}. Their work focused on confirming empirically that the mechanism for faults induced by voltage glitching is due to setup/hold time violations. They concluded that voltage glitches increased the propagation time of combinational logic which creates setup/hold time violations. Another connection we make is that while their paper wanted empirical evidence for a wide-held belief in how voltage glitches work we want a rigorous theoretical model for the same wide-held belief.

One level down from that is the paper by Djellid-Ouar, Cathebras and Bancel \cite{DCB} which concluded that D-flip-flops (and therefore most memory elements) were mostly immune to standard voltage attacks. Their paper analyzed bistable CMOS elements using the small signal model.

In addition to modeling voltage glitches, there has been many experiments in applying voltage fault injections to affect processor operations on complex SoCs. The aforementioned paper from Timmers, Spruyt, and Witteman \cite{TSW} details a bypass of a secure boot code integrity check on an ARM processor through voltage glitching. O'Flynn \cite{OFlynn2016FaultIU} used his crowbar method on a Raspberry Pi computer to modify the result of a running counter.

The PlayStation Vita was a hand-held gaming console released in 2012. It used a custom designed Samsung 45nm SoC \cite{anandtech}. The SoC includes a MeP architecture processor which we nicknamed ``F00D,'' that performs cryptographic tasks and serves as the boot processor. The boot ROM used by F00D is unmapped early in the boot process and is then unable to be read out through pure software means.

Our contribution will be in two separate domains. In section \ref{sec:model}, we will analyze the CMOS transistor behavior in order to understand \textit{when the combinational logic is most susceptible to voltage glitch induced faults}. Then, in section \ref{sec:vita} we will apply our understanding to perform a fault injection attack on the PlayStation Vita's SoC to gain early (boot time) execution control of F00D in order to dump the boot ROM.

\begin{table*}
\centering
\caption{Related works in modeling voltage glitches}
\begin{tabular}{lllll}
Paper            & Year & Level          & Applicability           & Results                                                     \\
\hline
TSW \cite{TSW}   & 2016 & Architectural  & ARM, any glitches       & Faults modeled by corrupted instructions                      \\
ZDCT \cite{ZDCT} & 2013 & Digital Logic  & Voltage/clock glitching & Fault caused by setup/hold violations                       \\
DCB \cite{DCB}   & 2006 & Gates/Elements & Voltage glitching       & D-flip-flops not susceptible to voltage glitches            \\
This Paper         & 2018 & Transistor     & Voltage glitching       & Effects are data dependent
\end{tabular}
\label{table:related}
\end{table*}

\section{CMOS VOLTAGE GLITCH MODEL}
\label{sec:model}

\begin{figure}
\centering
\begin{circuitikz} \draw
(0,2) node[ocirc] (in) {} node[anchor=east] {$V_{in}$}
(2,4) node[vcc] (vcc) {$V_{DD}$}
(2,3) node[pmos] (pmos) {}
(pmos.gate) node[anchor=east] {G}
(pmos.drain) node[anchor=east] {D}
(pmos.source) node[anchor=east] {S}
(4,2) node[inputarrow] (out) {$V_{out}$} 
(2,1) node[nmos] (nmos) {}
(nmos.gate) node[anchor=east] {G}
(nmos.drain) node[anchor=north] {}
(nmos.source) node[anchor=south] {}
(2,0) node[ground] (gnd) {} node[anchor=east] {$V_{SS}$}

(pmos.source) to (vcc)
(pmos.drain) to (nmos.drain)
(nmos.source) to (gnd)
(pmos.gate) to (nmos.gate)
(1,2) to (in)
(2,2) to (out)
(3,2) to[C,l=$C_L$] (3,0)
(3,0) to (2,0)
;
\end{circuitikz}
\caption{Standard CMOS inverter with load capacitance $C_L$}
\label{fig:cir1}
\end{figure}
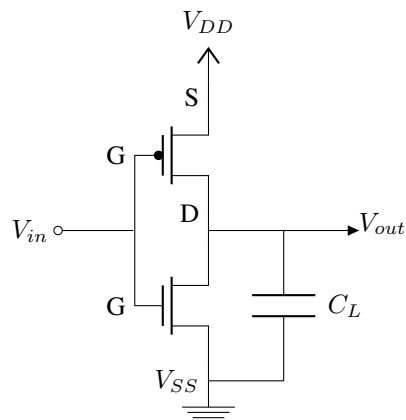

To analyze the CMOS behavior during a voltage glitch, we will only consider the voltages near the gate itself. This simplification will disregard everything that happens when the voltage pads on the IC is suddenly changed. For example, if a crowbar circuit is used to quickly short $V_{DD}$ to $V_{SS}$ for some amount of time, we do not actually observe a short at the MOSFET. Instead, the capacitance and the power-delivery-network of the circuit will create a ringing effect \cite{OFlynn2016FaultIU} that will be observed at the MOSFET. Our analysis will therefore only consider the duration and amplitude of these rings as the ``glitch'' and not the source of them. We note that a more in depth analysis can incorporate such external effects without affecting our understanding of what happens at the MOSFET.

\subsection{Notation}

We will use standard notation where applicable. For convenience, they are defined in table \ref{table:notation} along with other labels relevant in our analysis.

\begin{table}
\centering
\caption{Notations used}
\begin{tabular}{ll}
$V_{DD}$   & Supply voltage (normal operations)               \\
$V_{SS}$   & Ground voltage (typically $\SI{0}{\volt}$)                    \\
$V_{DD}'$  & Glitch supply voltage (typically $\sim\SI{0}{\volt}$)            \\
$V_{in}$   & CMOS input voltage                               \\
$V_{out}$  & CMOS output voltage                              \\
$C_L$      & Gate load capacitance (simplified)               \\
$V_{SG}$   & Voltage from PMOS source to PMOS gate            \\
$V_{TH}$   & PMOS threshold voltage                           \\
$V_{IL}$   & Switching threshold for input low                \\
$V_{IH}$   & Switching threshold for input high               \\
$R_{eqp}$  & PMOS equivalent resistance with source $V_{DD}$  \\
$R_{eqn}$  & NMOS equivalent resistance with source $V_{SS}$  \\
$R_{eqp}'$ & PMOS equivalent resistance with source $V_{DD}'$ \\
$t_{pHL}$  & Propagation delay for output going low           \\
$t_{pLH}$  & Propagation delay for output going high          \\
$\tau_A$   & Glitch start time                                \\
$\tau_B$   & Glitch end time                                  \\
$t_G$      & $\tau_B - \tau_A$                                \\
$t_{gHL}$  & Propagation delay for output going               \\
           & low during glitch (to be defined)                \\
$t_{gLH}$  & Propagation delay for output going               \\
           & high during glitch (to be defined)      
\end{tabular}
\label{table:notation}
\end{table}

\subsection{Motivation}

Consider a standard 1-input CMOS gate (an inverter, figure \ref{fig:cir1}). Let $V_{in}$ be the input voltage and $V_{out}$ be the output. We define a voltage glitch to be a span of time from $\tau_A$ to $\tau_B$ during which, we set $V_{DD} \gets V_{DD}'$. $V_{DD}'$ is the glitch voltage (and ideally $V_{DD}'=V_{SS}$). $t_G=\tau_B-\tau_A$ is the glitch width. $C_L$ represents the load capacitance and includes the gate parasitic capacitances, the wire capacitance, and the input capacitance of the next gate.


We will focus on the span of time between $\tau_A$ and $\tau_B$ when the voltage glitch happens. Looking at only one inverter, we see that in that span of time, the input could either toggle at least one time or not toggle at all. We will analyze the behavior of the inverter in both cases to see when the output value can be influenced by the glitch.

\subsection{Non-toggling}

First, consider the case when $V_{in}$ does not change during the voltage glitch. If the input is a logical `1' ($V_{in} \geq V_{IH}$) then the output would be at $V_{out}=V_{SS}$. Since $V_{SG}<V_{TH}$, the PMOS is off and there is no source for $V_{out}$ to change. Therefore the voltage glitch caused no change in behavior.

If the input is at logical `0' ($V_{in} \leq V_{IL}$), then before the voltage glitch, $V_{in}=V_{SS}$ and $V_{out}=V_{DD}$. When the glitch happens, the PMOS will be on, so we can use the equivalent resistance model defined in Chapter 5.4 of \cite{RC1} to analyze the dynamic behavior.

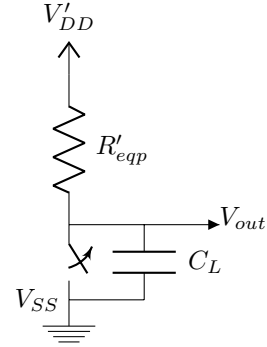
\begin{figure}
\centering
\begin{circuitikz} \draw
(0,3) node[vcc] (vdd) {$V_{DD}'$}
(0,0) node[ground] (gnd) {} node[anchor=east] {$V_{SS}$}
(2,1) node[inputarrow] (out) {$V_{out}$} 

(0,3) to[R,l=$R_{eqp}'$]
(0,1) to[ospst]
(0,0) {}
(0,1) to (out)
(1,1) to[C,l=$C_L$]
(1,0) to (gnd)
;
\end{circuitikz}
\caption{Equivalent circuit for PMOS in non-saturation.}
\label{fig:cir2}
\end{figure}

Using equation 5.17 in \cite{RC1}, we find the fall time to be 

\begin{equation}
t_{gHL}=\ln(2) R_{eqp} C_L
\end{equation}
where $R_{eqp}$ is the on-resistance of the PMOS at the glitch voltage $V_{DD}'$. Note that this is slightly different from the commonly used value $t_{pHL}$ (fall time of the inverter) because $t_{pHL}$ is defined with respect to $R_{eqn}$, the on-resistance of the NMOS.

The output goes to `0' for the duration of the glitch and then is be restored to `1' as $V_{DD}$ is restored to its pre-glitch value. This means that if the input is `0' at the start of the voltage glitch and does not toggle, then the output will be low from $\tau_A + t_{gHL}$ to $\tau_B + t_{pLH}$ (assuming $t_{gHL} \ll t_G$). Note this is similar to a static hazard, which can cause setup and hold time issues.

\subsection{Toggling}

Consider a transition from a logical `0' to `1' at the input during the voltage glitch (from $\tau_A$ to $\tau_B$). Since this turns off the PMOS, there should be no change in behavior at the output (compared to what happens if it toggles while there is no voltage glitch). However, if the transition is from a logical `1' to `0', then according to \cite{ZDCT}, the rise time of the inverter output,  $t_{gLH}$ increases as $V_{DD}'$ decreases. Additionally, if the glitch voltage $V_{DD}' < V_{TH}$, then the output will not go high for the duration of the voltage glitch ($t_{gLH}$ would be infinite). So this means the propagation delay increases as $V_{DD}'$ decreases up to $t_G$. This also causes setup and hold time issues.

\subsection{Results}

In both cases, we see that the delay introduced to a single gate by a voltage glitch is composed of a rise/fall time ($t_{gHL}$ or $t_{gLH}$ depending on the input value), the glitch width $t_G$, and finally the delay for when $V_{DD}$ is ``restored'' and the ``correct'' input must propagate to the output again ($t_{pLH}$). We can define $t_g$ to be the propagation delay of that inverter during a voltage glitch from the input to the output (we consider the value to be ``propagated'' only when the output has the correct value). From above, we see this can be broken into four cases.

\begin{equation}
\begin{aligned}
    t_{g,\text{non-toggle},0} & \geq 
\begin{cases}
    t_{G}+t_{pLH}, & \text{if } t_{gHL} > t_G \\
    t_{G}-t_{gHL}+t_{pLH}, & \text{otherwise}
\end{cases} \\
    t_{g,\text{non-toggle},1} & = 0 \\
    t_{g,\text{toggle},0\text{-to-}1} & \geq 
\begin{cases}
    t_{G}+t_{pLH}, & \text{if } t_{gLH} > t_G \\
    t_{gLH}, & \text{otherwise}
\end{cases} \\
    t_{g,\text{toggle},1\text{-to-}0} & = t_{pHL}
\end{aligned}
\end{equation}

Now, if we have a chain of $N$ inverters, we can find the total propogation through the chain considering that only the first inverter is affected by the voltage glitch (with delay $t_{g0}=(t_{g,\text{non-toggle},0}+t_{g,\text{toggle},0\text{-to-}1})/2$, the average delay of the two cases that are affected by a voltage glitch). Note that if two inverters in a chain are affected, we do not care about the status of the later inverter because the earlier one will already propagate its corrupted value and later corrected value down the chain.

\begin{equation}
    t_g(N) = t_{g0} + (N-1)\frac{t_{pLH}+t_{pHL}}{2}
\label{equ:a}
\end{equation}

In practice $t_{gHL}$, $t_{gLH}$, $t_{pHL}$, $t_{pLH}$ will all be very small compared to $t_G$\footnote{$t_{pHL}$ and $t_{pLH}$ are around the order of \SI{30}{\pico\second} for \SI{0.25}{\micro\metre} CMOS technology \cite{RC1}.}. Therefore we can simplify equation \ref{equ:a} to:

\begin{equation}
    t_g(N) \geq t_{G} + N \frac{t_{pLH}+t_{pHL}}{2}
\end{equation}

Note that the second part is just the CMOS propagation time with fanout 1. That means as long as $t_G$ is much greater than the rise/fall time of the output, the propagation delay is bounded by the CMOS delay plus the glitch width.

Of course in reality, the analysis gets a lot more complicated. First, the voltage glitch will not affect every CMOS at the same time. There will be a sort of ``propagation delay'' of the voltage change itself. Second, when we consider 2-input gates and higher, there could be a mixture of non-toggling and toggling behavior at each gate. Then of course, there are the non-linear capacitance that makes computing $C_L$ difficult.

However, there are some conclusions we can draw from this analysis.

\begin{itemize}
    \item Asynchronous circuits are most affected by voltage glitches due to the introduction of hazards.
    \item Synchronous circuits are not immune if the voltage glitch cause a setup/hold time violation.
    \item Critical paths can be extended if a voltage glitch happens at the right time (0-static or 1-toggling).
    \item Long critical paths are the best targets for voltage glitching (i.e: processor ALU).
\end{itemize}

\section{VITA GLITCHING}
\label{sec:vita}

The firmware and boot loader are found on an external eMMC storage, which has logical sectors 512 bytes wide. Upon boot, sector 0, the master boot record is read. The Vita's MBR is a custom format not used in any other device. The details of this MBR format are beyond the scope of this paper, but two fields are of importance. $\mathtt{bldr\_offset}$ is at MBR offset $\mathtt{0x30}$ and $\mathtt{bldr\_size}$ is at MBR offset $\mathtt{0x34}$. These two fields are both 4 bytes wide and both the offset and size are defined in number of sectors. They are used by the boot ROM to determine where the boot loader is located.

One thing we discovered early on (through trial and error) was that if $\mathtt{bldr\_size} > \mathtt{0xDE}$, then an assertion fails and the device is rebooted. Otherwise, the eMMC is read starting at the offset for $\mathtt{bldr\_size}$ blocks. Our hypothesis is that there is a fixed size buffer that the boot loader is read into which necessitates the size check. If we use a fault injection to bypass this check, we can introduce a buffer overflow vulnerability.

\subsection{Experimental Setup}

There were three main components to our setup. First, we needed a way to monitor the eMMC traffic and use it as a trigger for the voltage glitch. Second, we needed a way to perform the voltage glitch. Finally, we wished to automate the steps in order to find the optimal parameters for glitching.

Fortunately, the ChipWhisperer gave us an easy way to do all of this. The hardware has a MOSFET that performs the crowbar voltage glitch \cite{OFlynn2016FaultIU}. The open hardware design allowed us to implement a custom eMMC trigger for the MOSFET (see appendix \ref{app:emmc}). Finally, because the timing and duration of the voltage glitch is highly dependent on the power distribution network of the device \cite{OFlynn2016FaultIU}, it is difficult to compute the optimal timing parameters for a successful glitch on the size check. Instead, we exhaustively searched for the timing offset and width of the crowbar activation after being triggered that results in a successful fault injection.

Additionally, we made sure to synchronize the ChipWhisperer’s glitch module clock with the device’s external clock input. This way we can make sure the two devices are in phase and decrease the variance in finding working parameters.

\subsection{Parameters}

For a successful fault injection, we need to find parameters $f$, the clock frequency of the glitch module and the Vita’s external clock input, $N$, the number of cycles after seeing the eMMC trigger before activating the crowbar circuit and $M$, the number of cycles to hold the crowbar on before releasing it. Note that these parameters will determine the response of the ringing effect that will ultimately cause a fault in the size comparison\footnote{Many sources mention removing decoupling capacitors for better result without giving a detailed reason. We were able to get voltage glitches to work both with and without removing the decoupling capacitors. It is our belief that removing the decoupling capacitors changes the response of the ringing and therefore the parameters for a successful glitch. But in our case, it does not make it any more or less tractable.}.

We know from experimentation\footnote{Toggle GPIO and measure with the Rigol DS1054Z.} that F00D boot ROM runs at $f_{clk} = \frac{1}{9} f$ (where $f$ is the external clock input). A faster clock will increase the chance of a successful fault (due to timing violations). In practice, we cannot over-clock much past the default rate of 37MHz or we will run into non-voltage glitch related timing violations that prevent the circuit from working properly\footnote{This might be prevented with, for example, better cooling but we did not go down this avenue.}. However, in our case, because of the noisy design of our glitching setup, we picked $f=\SI{12}{\mega\hertz}$ since it was the fastest we were able to go without running into a variety of signal integrity issues.

For $N$ and $M$, we chose to brute force every possible value to find ones that worked.

\begin{figure}
\centering
\begin{tikzpicture}[node distance=2cm]

\node (start) [startstop] {Power On};
\node (clk) [io, below of=start] {Set clock to $f$};
\node (reset) [process, below of=clk] {Reset};
\node (readmbr) [process, below of=reset] {Wait for MBR read};
\node (waitn) [io, right of=start, xshift=2cm] {Wait $N$ cycles};
\node (activate) [process, below of=waitn] {Activate Crowbar Circuit};
\node (waitm) [io, below of=activate] {Wait $M$ cycles};
\node (deactivate) [process, below of=waitm] {Deactivate Crowbar Circuit};
\node (seeread) [decision, below of=readmbr, yshift=-0.5cm] {See packet read?};
\node (poweron) [decision, below of=seeread, yshift=-1cm] {Power still on?};
\node (success) [startstop, right of=seeread, xshift=2cm] {Success!};
\node (incm) [io, right of=poweron, xshift=2cm] {$M \gets M+1$};
\node (incn) [io, below of=incm] {$M \gets 0$, $N \gets N+1$};

\draw [arrow] (start) -- (clk);
\draw [arrow] (clk) -- (reset);
\draw [arrow] (reset) -- (readmbr);
\draw [arrow] (readmbr) -| ++(2,0) |- (waitn.west);
\draw [arrow] (waitn) -- (activate);
\draw [arrow] (activate) -- (waitm);
\draw [arrow] (waitm) -- (deactivate);
\draw [arrow] (deactivate) -| ++(-1.8,0) |- (seeread.north);
\draw [arrow] (seeread) -- node[anchor=south] {Yes} (success);
\draw [arrow] (seeread) -- node[anchor=east] {No} (poweron);
\draw [arrow] (poweron) -- node[anchor=south] {Yes} (incm);
\draw [arrow] (poweron) |- node[anchor=east] {No} (incn);
\draw [arrow] (incm) -- ++(2,0) -- ++(0,-3) -- ++(-8,0) -- ++(0,10.5) -- (reset.west);
\draw [arrow] (incn) -- ++(2,0) -- ++(0,-1) -- ++(-8,0) -- ++(0,10.5) -- (reset.west);

\end{tikzpicture}
\caption{Parameter search process}
\label{fig:flow}
\end{figure}
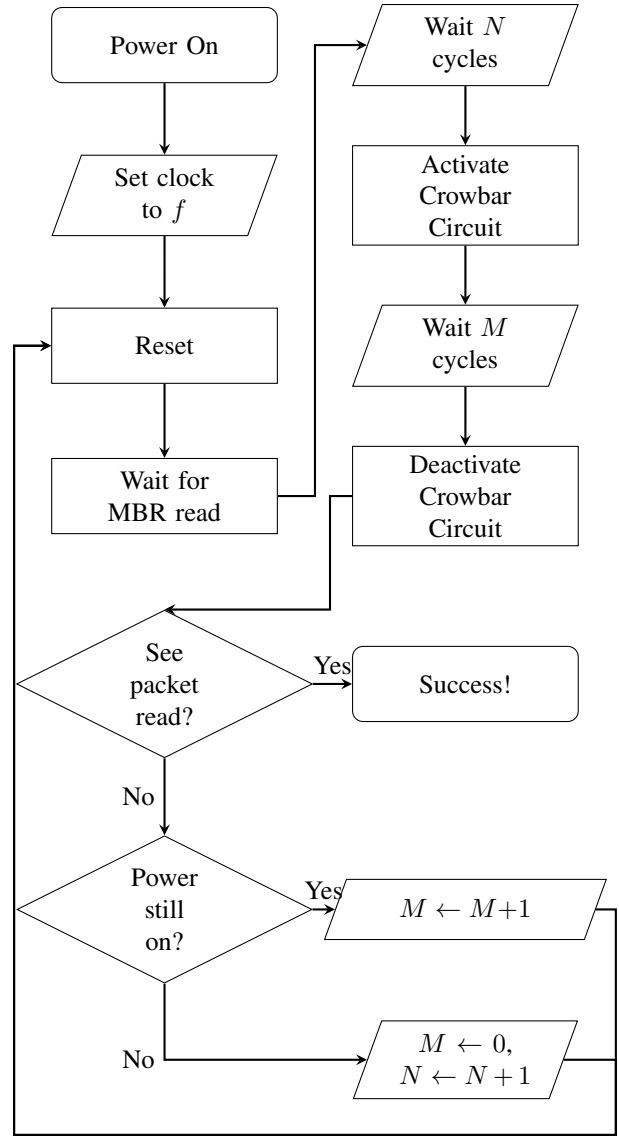

For the parameter search, we first used manual analysis to get a ballpark idea of when $\mathtt{bldr\_size}$ is being checked. We hypothesized that the check must happen after the response packet for the eMMC $\mathtt{READ\_SINGLE\_BLOCK}$ request for the MBR block. First we narrowed the window for the search to be the period of time between commands. We measured the amount of time from the response of $\mathtt{READ\_SINGLE\_BLOCK}$ packet (with a valid $\mathtt{bldr\_size}$) to the request of the $\mathtt{SEND\_STATUS}$\footnote{Defined as part of the eMMC read procedure \cite{emmc}.} packet using a Rigol DS1054Z oscilloscope. At $f_{clk}=\SI{12/9}{\mega\hertz}$, it took \SI{36}{\milli\second}\footnote{We discovered that due to a bug, the boot ROM spins the processor to wait for the $\mathtt{READ\_SINGLE\_BLOCK}$ request to complete. However, the smallest granularity of the spin time is about 1000 times the average amount of time it takes for the command to complete (as we observed on the DS1054Z). So to make things easier, our brute force actually ran backwards starting from the far end of the window.}. This yields an upper bound for $N$.

When an attempt fails with a particular $N$, $M$ pair, we observe one of the following behaviors:
\begin{itemize}
\item The device halts. Usually we only see this with very large M so it’s likely the CMOS is losing power and shutting off.
\item The device reboots and we observe the eMMC $\mathtt{GO\_IDLE\_STATE}$ packet. This is the most common observation\footnote{We believe this is due to the fault happening at the wrong place or not happening at all. Both of which would cause an assertion to fail. Originally we wanted to try some timing attack to differentiate the two cases. However we later found out that the designers actually anticipated this and masked any reboot triggered by an assertion fail to first spin for a random number of cycles determined by a TRNG to make it hard to determine ``which'' assertion fail caused the reboot.}.
\item The device goes into an unknown state and makes an unexpected request (and possibly restarts).
\item The device requests the first block of the bootloader (this is the success case).
\end{itemize}

We therefore only need to record the next packet seen after reading MBR (or time out) and check if it is the first block of the boot loader to indicate success. Figure \ref{fig:flow} summarizes the process.

\subsection{Results}

After writing a script with the ChipWhisperer API to try all possible $N$ and $M$ values (see appendix \ref{app:search}) and running it overnight, we found a successful case with the following parameters (see table \ref{table:results}). Due to the effects of wire capacitance and external capacitance, the parameters are highly specific to our setup and environment. However, after finding a valid $N$,$M$ pair, even with environmental variations, we can find another valid $N$,$M$ pair close by. If the equipment and target board and not moved or touched at all, we can reproduce the injected fault with the same $N$,$M$ for $> 80\%$ of the time.

\begin{table}
\centering
\caption{Range of working p}
\begin{tabular}{|l|l|l|l|}
\hline
Parameter & Units  & Min   & Max   \\ \hline
$f$         & \SI{}{\hertz}     & 12    & 12    \\ \hline
$N$         & cycles & 40800 & 40820 \\ \hline
$M$         & cycles & 45    & 55    \\ \hline
\end{tabular}
\label{table:results}
\end{table}

\subsection{Exploiting}

We managed to find the glitch parameters $N$ and $M$ that faults the $\mathtt{bldr\_size}$ check. However this only gives us a vulnerability. We still need to exploit it. Fortunately, this was made easy when we observed that when the glitch was successful, we see exactly $\mathtt{0xE2}$ blocks read if $\mathtt{bldr\_size} \geq \mathtt{0xE2}$. If $\mathtt{bldr\_size} < \mathtt{0xE2}$, then it will read $\mathtt{bldr\_size}$ blocks exactly (with a successful glitch). Therefore we guessed that we were overwriting the currently executed code and that guess was correct. With that, we launched a payload that dumped everything we can read through UART.

Unfortunately we then discovered the boot ROM does not seem to be mapped anywhere. It appears that hardware copies the contents of boot ROM to SRAM and the reset vector points directly to SRAM. That means the boot ROM is able to clean up parts of itself as it executes. To get the remaining parts (that were cleaned up), we had to glitch different parts of the boot ROM (running in SRAM) and gain code execution earlier and earlier on. Each time we dump more code, we gain more information and can develop more specific glitch targets. The details of these additional injected faults and their subsequent exploitation is beyond the scope of this paper.

\section{CONCLUSION}

By looking at how voltage glitches introduce timing violations into a digital circuit, we can find good snippets of code to glitch. Once a target is found, we can search for the right timing parameters for our crowbar circuit to cause a fault. We do an exhaustive search because it is difficult to predict how changing the parameters $N$ and $M$ actually affects the CMOS circuits. Finally, the injected fault introduces a software vulnerability that can be exploited to gain code execution. All of this can be done at a low cost thanks to the open hardware interface of the ChipWhisperer. With a custom script written for ChipWhisperer, we created a working attack on a security hardened consumer device.


\printbibliography

\appendices

\section{Code}

\subsection{Custom ChipWhisperer with eMMC Triggering}
\label{app:emmc}
\url{https://github.com/TeamMolecule/chipwhisperer}

\subsection{Parameters Search Script}
\label{app:search}
\url{https://github.com/TeamMolecule/petite-mort}

\end{document}